\documentclass[letterpaper]{jpconf}
\usepackage{graphicx}
\begin{document}
\title{Search for New Physics in NA62}

\author{Antonino Sergi\footnote{\uppercase{O}n behalf of the \uppercase{NA}62 \uppercase{C}ollaboration}}

\address{University of Perugia and INFN}

\begin{abstract}
The ratio $R_K=\frac{\Gamma(K\to e \nu_{e} (\gamma))}{\Gamma(K\to \mu \nu_{\mu} (\gamma))}$ 
of leptonic decay rates provides a golden probe for testing the structure of the weak 
interactions because it can be predicted with high accuracy within the Standard Model.
The aim of the NA62 experimental programme based on the 2007 data set is a measurement 
of $R_K$ reaching a new accuracy level better than $0.4\%$ \cite{evgueni}. 
To achieve this goal, data taking strategy allowing control over the systematic effects
and in particular precise background subtraction, was worked out, and a data sample 
of $\sim0.16\times10^6$ $K_{e2}$ candidates with just $\sim10\%$ background was collected.
The current status of the $K_{l2}$ analysis based on the dedicated NA62 data taking 
is summarized. The achieved precision of background subtraction, other systematic 
uncertainties, and prospects of the analysis are discussed.
Within the same scientific programme, NA62, in its second phase, will measure the branching ratio of the
very rare kaon decay $K^+ \rightarrow \pi^+ \nu \overline{\nu}$; the aim is to collect
$\mathcal{O}(100)$ events with 10\% background in two years data taking period.
The status of the project, the R\&D and future perspectives for the experiment
are discussed.
\end{abstract}

\section{Introduction}
\label{Intro}
In the Standard Model of Particle Physics (SM) sub-permille accuracy is reached
in the prediction of the ratios
$R_P=\frac{\Gamma(P\to e \nu_{e} (\gamma))}{\Gamma(P\to \mu \nu_{\mu} (\gamma))}=
\left(\frac{m_e}{m_\mu}\right)^2
\left(\frac{m_P^2-m_e^2}{m_P^2-m_\mu^2}\right)^2 (1 + \delta R_{\mathrm{QED}})$
of decay rates for the pseudoscalar $K$ and $\pi$ mesons. 
This excellent accuracy ($R_K=2.477\pm0.001 \times 10^{-5}$, $R_{\pi}=12.352\pm0.001 \times 10^{-5}$)
is due to cancellations of the hadronic uncertainties\cite{masiero, cirigliano} 
and the precise knowledge of the electromagnetic correction $\delta R_{\mathrm{QED}}$.

The factor $(m_e/m_\mu)^2$ accounts for helicity suppression and makes $R_P$ very sensitive 
to SM extensions involving Pseudo-Scalar currents and non-universal corrections, which are 
expected to induce variations of $R_P$ of the order $10^{-4} < \delta R_P/R_P < 10^{-2}$.
In particular SM extensions involving Lepton Flavour Violation (LFV) are currently not
ruled out by experiment.

The current world average $R^{\pi}_{K} = 2.45 \pm0.11 \times 10^{-5}$ (PDG) dates back to three
experiments performed in the 1970s, and has insufficient precision ($4.5\%$) for stringent SM tests. 
A series of recent preliminary results from NA48/2 and KLOE represents a significant improvement: 
Combining these results with the PDG value yields a precision of $1.3\%$\cite{antonelli}.

On the other hand, for what concerns the future programme, the $K^+ \rightarrow \pi^+ \nu \overline{\nu}$ decay 
is a flavor changing neutral current process which proceeds through box and electroweak penguin diagrams. 
The predicted values~\cite{Buras:2006gb,Isidori:2005xm,Mescia:2007kn} of the branching fraction are of 
the order of $8 \times 10^{-11}$. The hadronic matrix element can be parametrized in terms of 
the $K^+ \rightarrow \pi^0 e^+ \nu$ branching ratio, well known experimentally, reducing the theoretical 
uncertainty to about 5\% and thus making the $K^+ \rightarrow \pi^+ \nu \overline{\nu}$
decay a theoretically clean environment sensitive to new physics.

The existing measurement of the $K^+ \rightarrow \pi^+ \nu \overline{\nu}$ branching ratio, based on 7 observed events~\cite{Adler:2008zz}, 
is compatible with SM within errors, but a 10\% accuracy measurement of the branchig ratio is required to provide a decisive test of new physics scenarios. This is the goal of the proposed NA62 experiment at CERN SPS~\cite{Anelli:2005ju}, which aims to collect $\mathcal{O}(100)$  $K^+ \rightarrow \pi^+ \nu \overline{\nu}$ events in two years, keeping background contamination at the level of 10\%.
\section{Current experimental setup}
\label{Setup}
For the first phase of NA62 the experimental setup is largely based on NA48/2.
The beam line K12 at CERN SPS is capable of delivering simultaneous narrow momentum 
band $K^+$ and $K^-$ beams. A central momentum of 75 GeV/c was used during 2007 data taking.
Most of the data were taken with the $K+$ beam only because the muon sweeping system 
is more effective in rejecting $\mu^{+}$. Conversely, about $10\%$ of the data were recorded 
with $K^-$ beam only in order to measure the background induced by the beam halo via
$\mu \to e$ decays.
The following detectors, located downstream a vacuum decay volume ($80m$ in length), were
exploited: a magnetic spectrometer composed of four drift chambers (DCHs) and a spectrometric 
magnet (providing a track momentum resolution of $\delta p_{[GeV/c]}/p=0.47\% \oplus p\cdot0.02\%$);
a liquid krypton electromagnetic calorimeter (LKr) used for $\gamma$ detection and particle
identification; a scintillator hodoscope (HOD) used for timing measurement and to produce 
fast trigger signals.
A minimum bias trigger configuration was employed, resulting in relatively low trigger
purities, but high efficiencies for leptons with momentum $p>10GeV/c$.
%The $K_{e2}$ trigger condition consisted of a time
%coincidence of hits in the two planes of the HOD (the so called Q1 signal) with an energy
%deposition of at least 10 GeV in the LKr. The $K_{\mu2}$ trigger condition consisted of the Q1
%signal alone downscaled by a factor ranging from 50 to 150. Very loose requirements on
%the activity in the DCHs were additionally included into the trigger logic to enhance its
%purity against high multiplicity events.
The main data sample was taken during the four months of data taking in 2007.
About $4\times10^{6}$ SPS spills were recorded, corresponding to 300 TB of raw data. 
Additional data were collected in 2008 to address in detail some systematic uncertainties.
\section{Analysis}
\label{Analysis}
The analysis strategy is based on counting the numbers of reconstructed $K_{e2}$ and $K_{\mu2}$
candidates collected simultaneously, consequently the result does not rely on kaon flux
measurement and several systematic effects, such as parts of the trigger and detection
efficiencies, cancel in the ratio.
The analysis is performed independently in bins of momentum of the charged track ($15 GeV/c < p <65 GeV/c$), 
due to strong dependence of the backgrounds and acceptance on this variable.
The ratio $R_K$ is computed in each momentum bin after having subtracted the background events from the
$K_{e2}$ and $K_{\mu2}$ candidates. 
Correction factors are then applied in order to account for acceptances and for  
trigger, reconstruction and particle identification efficiencies.

Due to the topological similarity of $K_{e2}$ and $K_{\mu2}$ decays, a large part 
of the selection criteria is common for both decays. This leads to cancellations 
of the related systematic uncertainties in $R_K$.
Particle identification based on the ration of energy $E$ deposited in the 
calorimeter and the momentum measured by the spectrometer $p$
($0.95<E/p<1.10$ for electrons, $E/p<0.20$ for muons) and
kinematic identification obtained with the measurement of the missing invariant mass 
($M^2_{miss}=(p_K-p_{l})^2 < 0.01 (GeV/c)^2$ for electrons)
are exploited for separating $K_{e2}$ and $K_{\mu2}$ decays.
The average momentum of the incoming kaons is measured with $K\to3\pi$ decays and 
is used for $K_{l2}$ kinematical reconstruction ($M_{miss}$); a narrow kaon momentum 
spectrum ($\Delta p^{rms}_{K}/p_{K} \sim 2\%$) is used in order to minimize 
the corresponding contribution to the $M_{miss}$ resolution.
After selection, $0.16\times10^{6}$ $K_{e2}$ candidates are identified.
The present analysis is based on the study of a sub-sample ($40\%$) 
of the full data sample. 

The study of the background is the main topic of the analysis. Detailed studies
with data and MonteCarlo are performed in order to identify and measure the main sources 
of background. The following background contributions have been identified: (1) $K_{\mu2}$ events where the $\mu$ is mis-identified 
as an electron due to ``catastrophic'' bremsstrahlung in the LKr (the amount of background 
induced is $8.07\pm0.21\%$); (2) $K_{e2\gamma}$ events: by convention, the inner bremsstrahlung 
(IB) part of the radiative $K_{l2\gamma}$ process is included into $R_K$, while the structure 
dependent (SD) $K_{l2\gamma}$ process, which can not be accurately computed, is not. 
Its contribution has to be measured directly and is $1.29\pm0.32\%$; (3) Beam muon halo
with $\mu\to e$ decays: the $K_{l2}$ decay signature consists of a single reconstructed track, 
thus the background in $K_{l2}$ samples induced by the beam halo is an issue (amount 
of background induced is $1.23\pm0.07\%$).
Samples of reconstructed $K_{l2}$ candidates, of the charge not present in the beam, provide
measurements of the halo background. Thus the separate $K^+$ and $K^-$ beam only samples allow
for cross-subtraction of the halo background with precisions much better than statistical
uncertainty of the measurement; (4) minor sources of background are electrons from $K_{2\pi}$
and $K_{e3}$ decays (contributions at the level of $0.1\%$). 
\section{Future experimental setup for measuring $K^+ \rightarrow \pi^+ \nu \overline{\nu}$}
For the second phase of NA62, a highly intense (800 MHz rate) 75 GeV/c positively charged particle beam will be produced by directing a 400 GeV/c primary SPS proton beam  onto a Be target. The beam is positron free and is composed by 6\% of $K^+$ (unseparated beam), which will be  tagged by the existing differential Cerenkov counter called CEDAR~\cite{bovet}. The average rate seen by the downstream detectors will be 11 MHz, due to kaon decays and accidentals coming from the beam-line.

The beam spectrometer, measuring $p_K$ and the direction of the kaon track inside the decay volume, consists of three stations of a hybrid Si 
pixel detector. A time resolution of at least 200 ps per station is required to provide a suitable tag for the kaon track. The pion spectrometer, measuring $p_{\pi}$ and the pion direction, is made by low mass straw chambers placed in vacuum, in order to reduce multiple scattering. A 18 m long RICH located after the pion spectrometer, and filled with Ne at atmospheric pressure, is the core of particle identification: it will identify pions with momentum greater than 15 GeV/c and it will provide a time measurement of the downstream track with 100 ps time resolution. A hermetic combination of calorimeters, comprising the existing LKr calorimeter and covering up to 50 mrad, will be used to identify and veto the photons produced in kaon decays. In addition, a specific detector to veto muons tracks will be used downstream.

\section{Conclusions and prospects}
\label{Summary}
The measurements of $R_K$ in bins of track momentum ($p$) are found to be compatible and independent of $p$, which 
indicates good control over the main systematic effects. 
The total error is $0.6\%$ consisting of a statistical error of $0.43\%$ and 
systematic uncertainties of $0.4\%$ mainly due to background subtraction ($0.25\%$, $0.32\%$
and $0.1\%$, respectively, due to $K_{\mu2}$, $K_{e2\gamma}$ and beam halo background).
Further studies of second-order effects and minor background sources are currently underway 
in order to finalize the measurement of $R_K$ with the currently considered partial $40\%$ 
data sample. The total uncertainty of the partial result is $<0.7\%$ 
(below the $1\%$ level for the first time).

The whole NA62 data sample of $0.16$ millions $K_{e2}$ decay candidates, which is one order 
of magnitude larger than the existing world sample, allows pushing the statistical uncertainty to
a level below $0.3\%$. Significant improvements of the $K_{\mu2}$ background and beam halo
systematic uncertainties are possible thanks to the special data sets collected in 2008 and dedicated to
the Ke2 systematics studies. The uncertainty due to the $K_{e2\gamma}$ (SD) background is
expected to be decreased by a direct measurement of this process. The ultimate precision
of the measurement is expected to reach the $0.4\%$ level, meeting the goal declared in the
proposal\cite{Anelli:2005ju}.

The NA62 programme foresees a measurement of Br($K^+ \rightarrow \pi^+ \nu \overline{\nu}$) with a $\sim$10\% accuracy. The ultra-rare $K^+ \rightarrow \pi^+ \nu \overline{\nu}$ decay is a unique environment for new physics searches. The R\&D program is well advanced: during the 2007 run the Collaboration tested successfully a full length RICH counter prototype and a full length straw prototype in the actual vacuum tube.

\end{document}